\documentclass[journal,12pt,a4paper,onecolumn,draftcls]{IEEEtran}%
\usepackage{amsfonts}
\usepackage{amsmath}
\usepackage{amssymb}
\usepackage{graphicx}%
\providecommand{\U}[1]{\protect \rule{.1in}{.1in}}

\begin{document}

\title{Failsafe Mechanism Design of Multicopters Based on Supervisory Control Theory}
\author{Quan Quan*, Zhiyao Zhao, Liyong Lin, Peng Wang, Walter Murray Wonham,
\textit{Life Fellow}, \textit{IEEE}, and Kai-Yuan Cai\thanks{Q. Quan, Z. Zhao
P. Wang and K.-Y. Cai are with School of Automation Science and Electrical
Engineering,Beihang University, Beijing 100191, China (e-mail:
qq\_buaa@buaa.edu.cn, zzy\_buaa@buaa.edu.cn, wp2204@gmail.com,
kycai@buaa.edu.cn). L. Lin and W. M. Wonham are with the Department of
Electrical and Computer Engineering, University of Toronto, Toronto, ON M5S
3G4, Canada (e-mail: liyong.lin@utoronto.ca,wonham@control.utoronto.ca). The
corresponding author Q. Quan is also with the Department of Electrical and
Computer Engineering, University of Toronto as a visiting professor.}}
\maketitle

\begin{abstract}
In order to handle undesirable failures of a multicopter which occur in either
the pre-flight process or the in-flight process, a failsafe mechanism design
method based on supervisory control theory is proposed for the semi-autonomous
control mode \footnote{Most multicopters have two high-level control modes:
semi-autonomous control and full-autonomous control. Many open source
autopilots support both modes. The semi-autonomous control mode implies that
autopilots can be used to stabilize the attitude of multicopters, and also
they can help multicopters to hold the altitude and position. Under such a
mode, a multicopter will be still under the control of remote pilots. On the
other hand, the full-autonomous control mode implies that the multicopter can
follow a pre-programmed mission script stored in the autopilot which is made
up of navigation commands, and also can take off and land automatically. Under
such a mode, remote pilots on the ground only need to schedule the tasks
\cite{Quan2017}.}. Failsafe mechanism is a control logic that guides what
subsequent actions the multicopter should take, by taking account of real-time
information from guidance, attitude control, diagnosis, and other low-level
subsystems. In order to design a failsafe mechanism for multicopters, safety
issues of multicopters are introduced. Then, user requirements including
functional requirements and safety requirements are textually described, where
function requirements determine a general multicopter plant, and safety
requirements cover the failsafe measures dealing with the presented safety
issues. In order to model the user requirements by discrete-event systems,
several multicopter modes and events are defined. On this basis, the
multicopter plant and control specifications are modeled by automata. Then, a
supervisor is synthesized by monolithic supervisory control theory. In
addition, we present three examples to demonstrate the potential blocking
phenomenon due to inappropriate design of control specifications. Also, we
discuss the meaning of correctness and the properties of the obtained
supervisor. This makes the failsafe mechanism convincingly correct and
effective. Finally, based on the obtained supervisory controller generated by
TCT software, an implementation method suitable for multicopters is presented,
in which the supervisory controller is transformed into decision-making codes.

\end{abstract}


\section{Introduction}

\begin{figure}[h]
\begin{center}
{\normalsize \includegraphics[
scale=0.8]{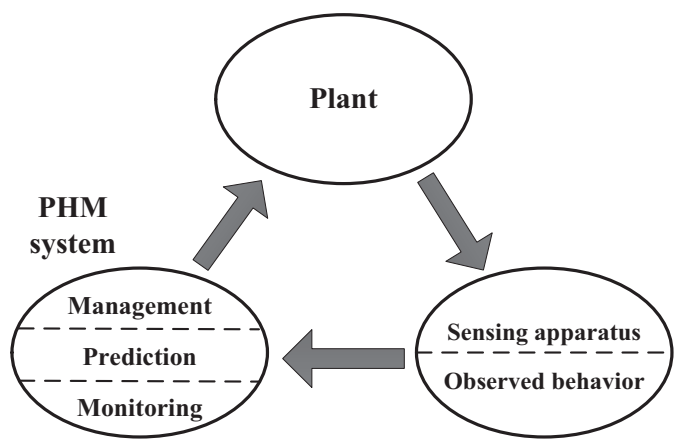} }
\end{center}
\caption{PHM framework}%
\label{PHM}%
\end{figure}

Multicopters are well-suited to a wide range of mission scenarios, such as
search and rescue \cite{Tomic2012}, \cite{Goodrich2008}, package delivery
\cite{Agha-mohammadi2014}, border patrol \cite{Girard2004}, military
surveillance \cite{Bethke2009} and agricultural production \cite{Huang2015}.
In either pre-flight process or in-flight process, multicopter failures cannot
be absolutely avoided. These failures may abort missions, crash multicopters,
and moreover, injure or even kill people. In order to handle undesirable
failures in industrial systems, a technique named Prognostics and Health
Management (PHM) is presented \cite{Kalgren2006}. As shown in Figure
\ref{PHM}, an integrated PHM system generally contains three levels:
monitoring, prediction and management \cite{Sheppard2009}. On the one hand,
the monitoring and prediction levels assess the quantitative health of the
studied system, where some quantitative indices are introduced to measure
system health, such as residuals \cite{Sun2012}-\cite{Wang2015}, data features
\cite{Henriquez2014}, \cite{Soualhi2014} and reliability-based indices
\cite{Xu2009}-\cite{Zhao2014}. On the other hand, the management level imports
the quantitative health results from the monitoring and prediction levels, and
then responds to meet qualitative safety or health requirements. In our
previous paper \cite{Zhao2017}, multicopter health is quantitatively evaluated
in the face of actuator failures. This paper studies a safety decision-making
logic by Supervisory Control Theory (SCT) to guarantee flight safety from a
qualitative perspective.

In the framework of multicopters, guidance, attitude control, PHM, and other
low-level subsystems work together under the coordination of a high-level
decision-making module \cite{Fisher2002}. In this module, a failsafe mechanism
is an important part. It is a control logic that receives information from all
subsystems to decide the best flight maneuver from a global perspective, and
send flight instructions to low-level subsystems \cite{Arnaiz2010}. However,
current academic literature covering failure-related topics of multicopters
mainly focuses on fault detection techniques \cite{Frangenberg2015}%
-\cite{Candido2014} and fault-tolerant control algorithms \cite{Falconi2016}%
-\cite{Raabe2013}, which belong to a study of low-level subsystems. For the
study of the high-level decision-making module, most research focuses on path
planning \cite{Bozhinoski2015}-\cite{Noriega2016} and obstacle avoidance
\cite{Orsag2015}, \cite{Nieuwenhuisen2013} of an individual multicopter, or
PHM-based mission allocation of a multicopter team \cite{Bethke2009},
\cite{Chen2014a}, \cite{Omidshafiei2016}. However, few studies have focused on
the failsafe mechanism design of an individual multicopter subject to multiple
potential failures. References \cite{Olson2014}, \cite{Harmsel2016} proposed
an emergency flight planning for an energy-constrained situation. Reference
\cite{Smet2015} proposed a failsafe design for an uncontrollable situation.
Reference \cite{Johry2016} designed multiple failsafe measures dealing with
different anomalies of unmanned aerial vehicles. Nevertheless, that research
only considers certain ad-hoc failsafe mechanisms for certain faults or
anomalies, and so far does not present a comprehensive failsafe mechanism for
a multicopter. In current autopilot products (for example, DJI autopilot
\cite{DJI Failsafe} and ArduPilot \cite{ArduPilot Failsafe}), there exist
comprehensive failsafe mechanisms to cope with communication, sensor and
battery failures, but such mechanisms are either proprietary, or can be
accessed only in part. Moreover, as far as the authors know, these failsafe
mechanisms are mainly developed and synthesized according to engineering
experience. As a result, such a development process lacks a theoretical
foundation; this will inevitably lead to man-made mistakes, logical bugs and
an incomplete treatment. Motivated by these, this paper first summarizes
safety issues and user requirements for multicopters in the semi-autonomous
control manner as comprehensively and systematically as possible, and then
uses SCT of Discrete-Event Systems (DES) to design a failsafe mechanism of multicopters.

\begin{figure}[h]
\begin{center}
{\normalsize \includegraphics[
scale=0.6]{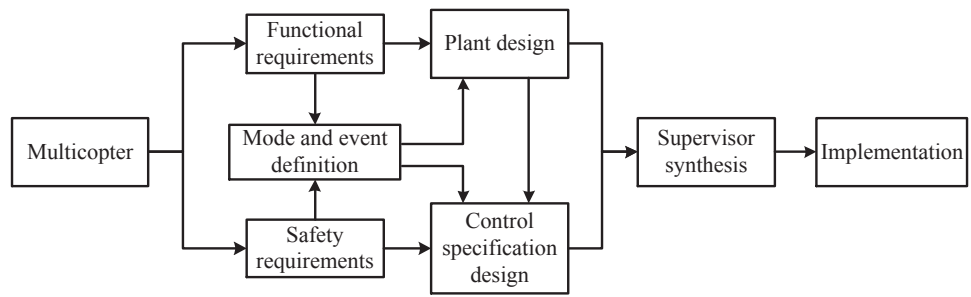} }
\end{center}
\caption{Solution procedure to design failsafe mechanism of multicopters}%
\label{sp}%
\end{figure}

SCT \cite{Wonham2009}, \cite{RW1987}, also known as Ramadge-Wonham (RW)
theory, is a method for synthesizing supervisors that restrict the behavior of
a plant such that as much as possible of the given control specifications are
fulfilled and never violated. Currently, SCT has been developed with a solid
theoretical foundation \cite{Cai2015}-\cite{Cai2014}, and it has been
successfully applied to practical systems such as flexible manufacturing
systems \cite{Leduc2006}-\cite{Hu2016}. Thus, this paper formalizes the
problem of failsafe mechanism design as a DES control problem. The solution
procedure is shown in Figure \ref{sp}. In order to obtain the expected
failsafe mechanism, the following steps are performed: 1) define related modes
and events by studying the user requirements (including functional and safety
requirements); 2) model the multicopter plant by transforming the functional
requirements to an automaton with defined modes and events; 3) analyze the
safety requirements by taking the defined modes and events into account, and
transform the safety requirements to automata as control specifications; 4)
synthesize the supervisor by \textit{TCT} software; 5) implementation the
failsafe mechanism based on the obtained supervisor.

The contributions of the paper mainly lie in two aspects.

\begin{itemize}
\item First, this paper introduces SCT into a new application area. The
proposed SCT based method in this paper is a scientific method with solid
theoretical foundation to design the failsafe mechanism of multicopters. In
the field of aircraft engineering, especially of multicopters and drones,
traditional design methods are based on engineering experience. The failsafe
mechanism obtained by these methods may be problematic (for example, the
failsafe mechanism may contain unintended deadlocks), especially when multiple
safety issues are taken into account. Compared to existing empirical design
methods, the proposed method can guarantee the correctness and effectiveness
of the obtained failsafe mechanism owing to the properties of supervisors.
This is an urgent need for multicopter designers and manufacturers.

\item Second, for the application of SCT, this paper emphasizes the modeling
process of the plant and control specifications with a practical application,
rather than developing a new theory of SCT. We believe this work is important
to both the development of SCT research and practical engineering, because SCT
is presented with complex mathematical terminology and theory which many
engineers may not understand. Motivated by this, this paper presents the
procedure of applying SCT to an engineering problem, from requirements
described textually, to specificarions in form of automata, then to a
synthesized supervisor and finally to implementation on a real-time flight
simulation platform of quadcopters developed by MATLAB. In addition, we
present three examples to demonstrate the potential blocking phenomenon due to
inappropriate design of control specifications. From the perspective of
practitioners, this paper can be a guide for engineers, who are not familiar
with SCT, to solve their own problems in their own projects by SCT and related software.
\end{itemize}

The remainder of this paper is organized as follows. Section II presents
preliminaries of SCT for the convenience of presenting the subsequent
sections. Section III lists some relevant safety issues of multicopters. Also,
user requirements including functional requirements and safety requirements
are textually described. In order to transform the user requirements to
automata, several multicopter modes and events are defined in Section IV. On
this basis, a detailed modeling process of the multicopter plant and control
specifications is presented in Section V, where functional requirements
determine a general multicopter plant, and safety requirements are modeled as
control specifications. Then, \textit{TCT} software is used to perform the
process of supervisor synthesis. Section VI illustrates three examples to
demonstrate some possible reasons leading to a problematic supervisor, and
gives a brief discussion about the scope of applications and properties of the
used method. Section VII shows an implementation process of the proposed
failsafe mechanism on the platform of MATLAB and FlightGear. Section VIII
presents our conclusion and suggests future research.

\section{Preliminaries of Supervisory Control Theory}

As SCT is well established, readers can refer to textbooks \cite{Wonham2009},
\cite{Fabian2004}, \cite{Cassandras2009} for detailed background and
knowledge. This section only reviews some basic concepts and notation.

In RW theory \cite{Wonham2009}, \cite{RW1987}, the formal structure of DES is
modeled by an automaton (generator)%
\begin{equation}
\mathbf{G}=\left(  Q,\Sigma,\delta,q_{0},Q_{m}\right)  \label{des1}%
\end{equation}
where $Q$ is the finite state set; $\Sigma$ is the finite event set (also
called an alphabet); $\delta:Q\times \Sigma \rightarrow Q$ is the (partial)
transition function; $q_{0}\in Q$ is the initial state; $Q_{m}\subseteq Q$ is
the subset of marker states. Let $\Sigma^{\ast}$ denote the set of all finite
strings, including the empty string $\epsilon$. In general, $\delta$ is
extended to $\delta:Q\times \Sigma^{\ast}\rightarrow Q$, and we write
$\delta \left(  q,s\right)  !$ to mean that $\delta \left(  q,s\right)  $ is
defined. The closed behavior of $\mathbf{G}$ is the language%
\[
L\left(  \mathbf{G}\right)  =\left \{  s\in \Sigma^{\ast}|\delta \left(
q,s\right)  !\right \}
\]
and the marked behavior is%
\[
L_{m}\left(  \mathbf{G}\right)  =\left \{  s\in L\left(  \mathbf{G}\right)
|\delta \left(  q_{0},s\right)  \in Q_{m}\right \}  \subseteq L\left(
\mathbf{G}\right)  .
\]
A string $s_{1}$ is a prefix of a string $s$, written $s_{1}\leqslant s$, if
there exists $s_{2}$ such that $s_{1}s_{2}=s$. The prefix closure of
$L_{m}\left(  \mathbf{G}\right)  $ is $\overline{L_{m}\left(  \mathbf{G}%
\right)  }:=\left \{  s_{1}\in \Sigma^{\ast}|\left(  \exists s\in L_{m}\left(
\mathbf{G}\right)  \right)  \text{ }s_{1}\leqslant s\right \}  $. We say that
$\mathbf{G}$\ is nonblocking if $\overline{L_{m}\left(  \mathbf{G}\right)
}=L\left(  \mathbf{G}\right)  $. The three equivalent meanings of
\textquotedblleft nonblocking\textquotedblright \ are 1) the system can always
reach a marker state from every reachable state; 2) every string in the closed
behavior can be extended to a string in the marked behavior; 3) every
physically possible execution can be extended to completing distinguished tasks.

The usual way to combine several automata into a single, more complex
automaton is called \emph{synchronous product}. For two automata
$\mathbf{G}_{i}=\left(  Q_{i},\Sigma_{i},\delta_{i},q_{0,i},Q_{m,i}\right)
,i=1,2$, the synchronous product $\mathbf{G}=\left(  Q,\Sigma,\delta
,q_{0},Q_{m}\right)  $ of $\mathbf{G}_{1}$ and $\mathbf{G}_{2}$, denoted by
$\mathbf{G}_{1}\Vert \mathbf{G}_{2}$, is constructed to have marked behavior
$L_{m}\left(  \mathbf{G}\right)  =L_{m}\left(  \mathbf{G}_{1}\right)  \Vert
L_{m}\left(  \mathbf{G}_{2}\right)  $ and closed behavior $L\left(
\mathbf{G}\right)  =L\left(  \mathbf{G}_{1}\right)  \Vert L\left(
\mathbf{G}_{2}\right)  $ \cite[Chapter 3.3]{Wonham2009}. The synchronous
product of more than two automata can be constructed similarly.

For a practical system, the plant can be modeled as an automaton $\mathbf{G}$.
The desired behavior of the controlled system is determined by a control
specification, also modeled as an automaton $\mathbf{E}$. Both the plant
$\mathbf{G}$\ and the control specification $\mathbf{E}$\ may be the
synchronous product of many smaller components.

For supervisory control, the alphabet $\Sigma$ is partitioned as%
\[
\Sigma=\Sigma_{c}\dot{\cup}\Sigma_{u}%
\]
where $\Sigma_{c}$ is the subset of controllable events that can be disabled
by an external supervisor, and $\Sigma_{u}$ is the subset of uncontrollable
events that cannot directly be prevented from occurring. Here $\Sigma_{c}$ and
$\Sigma_{u}$ are disjoint subsets. A supervisory controller (supervisor)
forces the plant to respect the control specification by disabling certain
controllable events that are originally able to occur in the plant.

To synthesize a satisfactory supervisor, SCT provides a formal method for
theoretically solving the typical supervisory control problem \cite{Feng2009}:
Given a plant $\mathbf{G}$\ over alphabet $\Sigma=\Sigma_{c}\dot{\cup}%
\Sigma_{u}$ and control specification $\mathbf{E}$, find a maximally
permissive supervisor $\mathbf{S}$ such that the controlled system
$\mathbf{S/G}$ is non-blocking and meets the control specification
$\mathbf{E}$. That is $\mathbf{S}$ satisfies%
\begin{equation}%
\begin{array}
[c]{l}%
L_{m}\left(  \mathbf{S}\right)  =\sup \mathcal{C}\left(  \mathbf{E\cap}%
L_{m}\left(  \mathbf{G}\right)  \right)  \subseteq L_{m}\left(  \mathbf{G}%
\right) \\
\overline{L_{m}\left(  \mathbf{S}\right)  }=L\left(  \mathbf{S}\right)
\end{array}
\label{des2}%
\end{equation}
where $\sup \mathcal{C}\left(  L\right)  $ means the supremal controllable
sublanguage of $L$. Equation (\ref{des2}) means that the supervisor
$\mathbf{S}$\ never violates the control specification $\mathbf{E}$. Here,
$\mathbf{S}$ is a monolithic (namely fully centralized) supervisor
\cite[Chapter 4.6]{Wonham2009}. If there exist several control specifications,
the supervisor can be also designed in a decentralized framework.
Decentralized supervisory control assigns a separate specialized supervisor to
satisfy each control specification $\mathbf{E}_{j}$. For each control
specification $\mathbf{E}_{j}$, a decentralized supervisor $\mathbf{S}_{j}$ is
computed in the same way as for a monolithic supervisor. Then, all the
decentralized supervisors work together to meet the control specification
$\mathbf{E}=\mathbf{E}_{1}\Vert \mathbf{E}_{2}\Vert \cdots$. Here, if the
synthesized supervisors are blocking, a coordinator is required to make the
supervisors nonblocking. The main advantage of the decentralized supervisory
control framework is that the synthesized supervisors are relatively
small-scale, and are easier to understand, maintain and change.

Related algorithms in DES and SCT can be performed on software platforms such
as \textit{TCT} software \cite{Wonham2009}, \textit{Supremica}
\cite{Akesson(2003)} and Discrete Event Control Kit written in MATLAB
\cite{Zad(2003)}.

\section{Safety Issues and User Requirements}

This section lists some relevant safety issues of multicopters. Also, user
requirements including functional requirements and safety requirements are
textually described.

\subsection{Safety issues}

Major types of multicopter failures that may cause accidents will be
introduced. Here, three types of failures are considered, including
communication breakdown, sensor failure and propulsion system anomaly.

\begin{itemize}
\item \emph{Communication breakdown}. Communication breakdown mainly refers to
a contact anomaly between the Remote Controller (RC) transmitter and the
multicopter, or between the Ground Control Station (GCS) and the multicopter.
In this paper, for simplicity, only RC is considered.

\item \emph{Sensor failure. }Sensor failure mainly implies that a sensor on
the multicopter cannot accurately measure related variables, or cannot work
properly. This paper considers the sensor failures including barometer
failure, compass failure, GPS failure, Inertial Navigation System (INS) failure.

\item \emph{Propulsion system anomaly. }Propulsion system anomaly mainly
refers to battery failure and propulsor failure caused by Electronic Speed
Controllers (ESCs), motors or propellers.
\end{itemize}

More information about safety issues can be found in the book \cite{Quan2017}.

\subsection{User requirements}

From the commercial perspective of customers and users, a multicopter product
is required to have general functions as a rotorcraft, and also be capable of
coping with the relevant safety issues. Thus, functional requirements and
safety requirements are listed in Tables 1-4, respectively. They are
summarized by referring the material from \cite{ArduPilot Failsafe} and the
authors' knowledge and engineering experience.

\subsubsection{Functional requirements}

The following functional requirements describe what behavior the multicopter
is able to perform.

\begin{center}
Table 1. Functional requirements%
\[%
\begin{tabular}
[c]{|c|c|}\hline \hline
{\small Name} & {\small Description}\\ \hline \hline
\multicolumn{1}{|l|}{{\small FR1}} & \multicolumn{1}{|l|}{{\small The remote
pilot can arm\footnotemark the multicopter by the RC transmitter and then
allow it to take off.}}\\ \hline
\multicolumn{1}{|l|}{{\small FR2}} & \multicolumn{1}{|l|}{{\small After taking
off, the remote pilot can manually switch the multicopter to fly normally,
return to the}}\\
\multicolumn{1}{|l|}{} & \multicolumn{1}{|l|}{{\small base or land
automatically by the RC transmitter.}}\\ \hline
\multicolumn{1}{|l|}{{\small FR3}} & \multicolumn{1}{|l|}{{\small The remote
pilot can manually control the multicopter to land and disarm it by the RC
transmitter.}}\\ \hline
\multicolumn{1}{|l|}{{\small FR4}} & \multicolumn{1}{|l|}{{\small When the
multicopter is flying, the multicopter can realize spot hover, altitude-hold
hover and attitude}}\\
\multicolumn{1}{|l|}{} & \multicolumn{1}{|l|}{{\small self-stabilization.}%
}\\ \hline
\multicolumn{1}{|l|}{{\small FR5}} & \multicolumn{1}{|l|}{{\small When the
multicopter is flying, the multicopter can automatically switch to returning
to the base or land.}}\\ \hline \hline
\end{tabular}
\  \  \  \  \
\]
\footnotetext{Arm is the instruction that the propellers of the multicopter be
unlocked; in this case, the multicopter can take off. Correspondingly, disarm
is the instruction that the propellers of the multicopter be locked; in this
case, the multicopter cannot take off.}
\end{center}

\subsubsection{Safety requirements}

The safety requirements restrict what action the user wants the multicopter to
perform under specific situations when it is on the ground, in flight, or in
process of returning and landing.

\begin{center}
Table 2. Safety requirements on ground%
\[%
\begin{tabular}
[c]{|c|c|}\hline \hline
{\small Name} & {\small Description}\\ \hline \hline
\multicolumn{1}{|l|}{{\small SR1}} & \multicolumn{1}{|l|}{{\small When the
remote pilot tries to arm the multicopter, if the INS and propulsors are both
healthy,}}\\
\multicolumn{1}{|l|}{} & \multicolumn{1}{|l|}{{\small the connection to RC
transmitter is normal, and the battery's capacity is adequate, then the}}\\
\multicolumn{1}{|l|}{} & \multicolumn{1}{|l|}{{\small multicopter can be
successfully armed and take off. Otherwise, the multicopter cannot be armed.}%
}\\ \hline \hline
\end{tabular}
\]

Table 3. Safety requirements in flight%
\[%
\begin{tabular}
[c]{|c|c|}\hline \hline
{\small Name} & {\small Description}\\ \hline \hline
\multicolumn{1}{|l|}{{\small SR2}} & \multicolumn{1}{|l|}{{\small If the
multicopter is already on the ground, the multicopter can be manually disarmed
by the }}\\
\multicolumn{1}{|l|}{} & \multicolumn{1}{|l|}{{\small RC transmitter, or
automatically disarmed if no instruction is sent to the multicopter by}}\\
\multicolumn{1}{|l|}{} & \multicolumn{1}{|l|}{{\small the RC transmitter.}%
}\\ \hline
\multicolumn{1}{|l|}{{\small SR3}} & \multicolumn{1}{|l|}{{\small When the
multicopter is flying, if the GPS or compass is unhealthy, the multicopter can
only}}\\
\multicolumn{1}{|l|}{} & \multicolumn{1}{|l|}{{\small realize altitude-hold
hover rather than spot hover. If the barometer is unhealthy, the multicopter}%
}\\
\multicolumn{1}{|l|}{} & \multicolumn{1}{|l|}{{\small can only realize
attitude self-stabilization. If the corresponding components are recovered,
the}}\\
\multicolumn{1}{|l|}{} & \multicolumn{1}{|l|}{{\small multicopter should
switch to an advanced hover status.}}\\ \hline
\multicolumn{1}{|l|}{{\small SR4}} & \multicolumn{1}{|l|}{{\small When the
multicopter is flying and the connection to the RC transmitter becomes
abnormal, if the}}\\
\multicolumn{1}{|l|}{} & \multicolumn{1}{|l|}{{\small INS, GPS, barometer,
compass and propulsors are all healthy, the multicopter should switch to}}\\
\multicolumn{1}{|l|}{} & \multicolumn{1}{|l|}{{\small returning to the base.
Otherwise, the multicopter should switch to landing.}}\\ \hline
\multicolumn{1}{|l|}{{\small SR5}} & \multicolumn{1}{|l|}{{\small When the
multicopter is flying, if the battery's capacity becomes inadequate but the
multicopter is}}\\
\multicolumn{1}{|l|}{} & \multicolumn{1}{|l|}{{\small able to return to the
base, then the multicopter should switch to returning to the base; if the
battery's}}\\
\multicolumn{1}{|l|}{} & \multicolumn{1}{|l|}{{\small capacity becomes
inadequate and unable to return, then the multicopter should switch to
landing.}}\\ \hline
\multicolumn{1}{|l|}{{\small SR6}} & \multicolumn{1}{|l|}{{\small When the
multicopter is flying, if the INS or propulsors are unhealthy, the
multicopter}}\\
\multicolumn{1}{|l|}{} & \multicolumn{1}{|l|}{{\small should automatically
switch to landing.}}\\ \hline
\multicolumn{1}{|l|}{{\small SR7}} & \multicolumn{1}{|l|}{{\small When the
multicopter is flying, the multicopter can be manually switched to returning
to the base by}}\\
\multicolumn{1}{|l|}{} & \multicolumn{1}{|l|}{{\small the RC transmitter. This
switch requires that the INS, GPS, barometer, compass, propulsors are}}\\
\multicolumn{1}{|l|}{} & \multicolumn{1}{|l|}{{\small all healthy, and the
battery's capacity is able to support the multicopter to return to the base.}%
}\\
\multicolumn{1}{|l|}{} & \multicolumn{1}{|l|}{{\small Otherwise, the switch
cannot occur.}}\\ \hline
\multicolumn{1}{|l|}{{\small SR8}} & \multicolumn{1}{|l|}{{\small When the
multicopter is flying, the multicopter can be manually switched to
automatically}}\\
\multicolumn{1}{|l|}{} & \multicolumn{1}{|l|}{{\small landing by the RC
transmitter.}}\\ \hline \hline
\end{tabular}
\
\]

Table 4. Safety requirements on returning and landing%
\[%
\begin{tabular}
[c]{|l|l|}\hline \hline
{\small Name} & {\small Description}\\ \hline \hline
{\small SR9} & {\small When the multicopter is in the process of returning to
the base, the multicopter can be}\\
& {\small manually switched to normal flight or landing by the RC
transmitter.}\\ \hline
{\small SR10} & {\small When the multicopter is in the process of returning to
the base, if the distance to}\\
& {\small the base is less than a given threshold, the multicopter should
switch to landing; if the}\\
& {\small battery's capacity becomes inadequate and unable to return to the
base, the multicopter}\\
& {\small should switch to landing; if the INS, GPS, barometer, compass or
propulsors are unhealthy,}\\
& {\small the multicopter should switch to landing.}\\ \hline
{\small SR11} & {\small When the multicopter is in the process of landing, the
multicopter can be manually switched}\\
& {\small to normal flight by the RC transmitter. This switch requires that
the INS and propulsors are}\\
& {\small both healthy, the connection to the RC transmitter is normal, and
the battery's capacity is}\\
& {\small adequate. Otherwise, the switch cannot occur.}\\ \hline
{\small SR12} & {\small When the multicopter is in the process of landing, the
multicopter can be manually switched}\\
& {\small to returning to the base by the RC transmitter. This switch requires
that the INS, GPS, barometer,}\\
& {\small compass, propulsors are all healthy, the battery's capacity is able
to support the multicopter to}\\
& {\small return to the base, and the distance to the base is not less than a
given threshold.}\\
& {\small Otherwise, the switch cannot occur.}\\ \hline
{\small SR13} & {\small When the multicopter is in the process of landing, if
the multicopter's altitude is lower}\\
& {\small than a given threshold, or the multicopter's throttle is less than a
given threshold }\\
& {\small over a time horizon, the multicopter can be automatically
disarmed.}\\ \hline \hline
\end{tabular}
\
\]

\end{center}

\section{Multicopter Mode and Event Definition}

In order to transform the user requirements to automata, several multicopter
modes and events are defined in this section.

\subsection{Multicopter mode}

Referring to \cite{ArduPilot Failsafe}, the whole process from taking off to
landing of multicopters is divided into eight multicopter modes. They form the
basis of the failsafe mechanism.

\begin{itemize}
\item POWER OFF MODE. This mode implies that a multicopter is out of power. In
this mode, the remote pilot can (possibly) disassemble, maintain and replace
the hardware of a multicopter.

\item STANDBY MODE. When a multicopter is connected to the power module, it
enters a pre-flight status. In this mode, the multicopter is not armed, and
the remote pilot can arm the multicopter manually. Afterwards, the multicopter
will perform a safety check and then transit to the next mode according to the
results of the safety check.

\item GROUND-ERROR MODE. This mode indicates that the multicopter has a safety
problem. In this mode, the buzzer will turn on an alarm to alert the remote
pilot that there exist errors in the multicopter.

\item LOITER MODE. Under this mode, the remote pilot can use the control
sticks of the RC transmitter to control the multicopter. Horizontal location
can be adjusted by the roll and pitch control sticks. When the remote pilot
releases the control sticks, the multicopter will slow to a stop. Altitude can
be controlled by the throttle control stick. The heading can be set with the
yaw control stick. When the remote pilot releases the roll, pitch and yaw
control sticks and pushes the throttle control stick to the mid-throttle
deadzone, the multicopter will automatically maintain the current location,
heading and altitude.

\item ALTITUDE-HOLD MODE. Under this mode, a multicopter maintains a
consistent altitude while allowing roll, pitch and yaw to be controlled
normally. When the throttle control stick is in the mid-throttle deadzone, the
throttle is automatically controlled to maintain the current altitude and the
attitude is also stabilized but the horizontal position drift will occur. The
remote pilot will need to regularly give roll and pitch commands to keep the
multicopter in place. When the throttle control stick goes outside the
mid-throttle deadzone, the multicopter will descend or climb depending upon
the deflection of the control stick.

\item STABILIZE MODE. This mode allows a remote pilot to fly the multicopter
manually, but self-levels the roll and pitch axes. When the remote pilot
releases the roll and pitch control sticks, the multicopter\ automatically
stabilizes its attitude but position drift may occur. During this process, the
remote pilot will need to regularly give roll, pitch and throttle commands to
keep the multicopter in place as it might be pushed around by wind.

\item RETURN-TO-LAUNCH (RTL) MODE. Under this mode, the multicopter will
return to the base location from the current position, and hover there.

\item AUTOMATIC-LANDING (AL) MODE. In this mode, the multicopter realizes
automatic landing by adjusting the throttle according to the estimated height
\footnote{Even if the barometer fails, the height estimation is acceptable
within a short time. Similarly, the other estimates generated by filters could
continue to be used for a short time, even if related sensors fail.}.
\end{itemize}

\subsection{Event definition}

Three types of events are defined here: Manual Input Events (MIEs), Mode
Control Events (MCEs) and Automatic Trigger Events (ATEs). The failsafe
mechanism detects the occurrence of MIEs and ATEs, and uses MCEs to decide
which mode the multicopter should stay in or switch to. Here, MIEs and MCEs
are controllable, while ATEs are uncontrollable in the sense of SCT.

\subsubsection{MIEs}

MIEs are instructions from the remote pilot sent through the RC transmitter.
This part defines eight MIEs as shown in Table 5.

\begin{center}
Table 5. MIE definition%
\[%
\begin{tabular}
[c]{|c|l|}\hline \hline
{\small Name} & {\small Description}\\ \hline \hline
{\small \textit{MIE1}} & {\small Turn on the power.}\\ \hline
{\small \textit{MIE2}} & {\small Turn off the power}\\ \hline
{\small \textit{MIE3}} & {\small Execute arm action. This action is realized
by manipulating the sticks of the RC transmitter.}\\ \hline
{\small \textit{MIE4}} & {\small Execute disarm action.}\\ \hline
{\small \textit{MIE5}} & {\small Other actions manipulated by the sticks of
the RC transmitter. These actions correspond to}\\
& {\small normal operations by the remote pilot. Here, no manipulation on the
sticks is also inclusive.}\\ \hline
{\small \textit{MIE6}} & {\small Switch to normal flight. In normal flight,
the multicopter can be in either LOITER MODE,}\\
& {\small ALTITUDE-HOLD MODE or STABILIZE MODE.}\\ \hline
{\small \textit{MIE7}} & {\small Switch to RTL MODE.}\\ \hline
{\small \textit{MIE8}} & {\small Switch to AL MODE.}\\ \hline \hline
\end{tabular}
\  \
\]

\end{center}

\begin{figure}[h]
\begin{center}
\includegraphics[
scale=1]{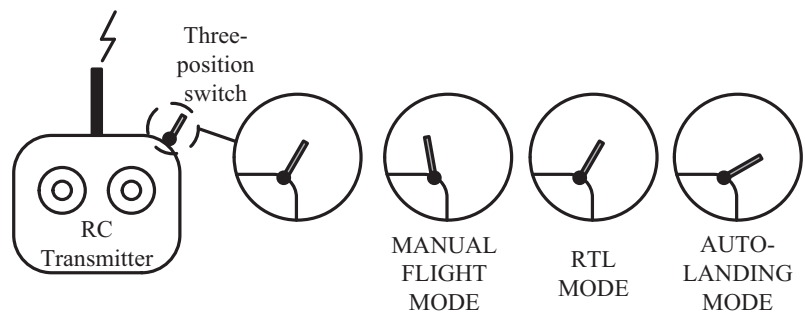}
\end{center}
\caption{Flight mode switch}%
\label{flight_mode_switch}%
\end{figure}

Here, \textit{MIE6}, \textit{MIE7} and \textit{MIE8} are realized by a
three-position switch (namely the flight mode switch) on the RC transmitter as
shown in Figure \ref{flight_mode_switch}.

\subsubsection{MCEs}

MCEs are instructions from multicopter's autopilot. As shown in Table 6, these
events will control the multicopter to switch to a specified mode.

\begin{center}
Table 6. MCE definition%
\[%
\begin{tabular}
[c]{|c|c|}\hline \hline
{\small Name} & {\small Description}\\ \hline \hline
{\small \textit{MCE1}} & \multicolumn{1}{|l|}{{\small Multicopter switched to
POWER OFF MODE.}}\\ \hline
{\small \textit{MCE2}} & \multicolumn{1}{|l|}{{\small Multicopter switched to
STANDBY MODE.}}\\ \hline
{\small \textit{MCE3}} & \multicolumn{1}{|l|}{{\small Multicopter switched to
GROUND-ERROR MODE.}}\\ \hline
{\small \textit{MCE4}} & \multicolumn{1}{|l|}{{\small Multicopter switched to
LOITER MODE.}}\\ \hline
{\small \textit{MCE5}} & \multicolumn{1}{|l|}{{\small Multicopter switched to
ALTITUDE-HOLD MODE.}}\\ \hline
{\small \textit{MCE6}} & \multicolumn{1}{|l|}{{\small Multicopter switched to
STABILIZE MODE.}}\\ \hline
{\small \textit{MCE7}} & \multicolumn{1}{|l|}{{\small Multicopter switched to
RTL MODE.}}\\ \hline
{\small \textit{MCE8}} & \multicolumn{1}{|l|}{{\small Multicopter switched to
AL MODE.}}\\ \hline \hline
\end{tabular}
\
\]

\end{center}

\subsubsection{ATEs}

ATEs are independent of the remote pilot's operations. As shown in Table 7,
these events contain the health check results of onboard equipment and sensor
measurements of the multicopter status.

\begin{center}
Table 7. ATE definition%
\[%
\begin{tabular}
[c]{|c|c|}\hline \hline
{\small Name} & {\small Description}\\ \hline \hline
{\small \textit{ATE1}} & \multicolumn{1}{|l|}{{\small The check result of INS
is healthy.}}\\ \hline
{\small \textit{ATE2}} & \multicolumn{1}{|l|}{{\small The check result of INS
is unhealthy.}}\\ \hline
{\small \textit{ATE3}} & \multicolumn{1}{|l|}{{\small The check result of GPS
is healthy.}}\\ \hline
{\small \textit{ATE4}} & \multicolumn{1}{|l|}{{\small The check result of GPS
is unhealthy.}}\\ \hline
{\small \textit{ATE5}} & \multicolumn{1}{|l|}{{\small The check result of
barometer is healthy.}}\\ \hline
{\small \textit{ATE6}} & \multicolumn{1}{|l|}{{\small The check result of
barometer is unhealthy.}}\\ \hline
{\small \textit{ATE7}} & \multicolumn{1}{|l|}{{\small The check result of
compass is healthy.}}\\ \hline
{\small \textit{ATE8}} & \multicolumn{1}{|l|}{{\small The check result of
compass is unhealthy.}}\\ \hline
{\small \textit{ATE9}} & \multicolumn{1}{|l|}{{\small The check result of
propulsors is healthy.}}\\ \hline
{\small \textit{ATE10}} & \multicolumn{1}{|l|}{{\small The check result of
propulsors is unhealthy.}}\\ \hline
{\small \textit{ATE11}} & \multicolumn{1}{|l|}{{\small The check result of
connection to the RC transmitter is normal.}}\\ \hline
{\small \textit{ATE12}} & \multicolumn{1}{|l|}{{\small The check result of
connection to the RC transmitter is abnormal.}}\\ \hline
{\small \textit{ATE13}} & \multicolumn{1}{|l|}{{\small The measured battery's
capacity is adequate.}}\\ \hline
{\small \textit{ATE14}} & \multicolumn{1}{|l|}{{\small The measured battery's
capacity is inadequate, able to RTL.}}\\ \hline
{\small \textit{ATE15}} & \multicolumn{1}{|l|}{{\small The measured battery's
capacity is inadequate, unable to RTL.}}\\ \hline
{\small \textit{ATE16}} & \multicolumn{1}{|l|}{{\small The measured
multicopter's altitude is lower than a given threshold.}}\\ \hline
{\small \textit{ATE17}} & \multicolumn{1}{|l|}{{\small The measured
multicopter's altitude is not lower than a given threshold.}}\\ \hline
{\small \textit{ATE18}} & \multicolumn{1}{|l|}{{\small The measured
multicopter's distance from the base is less than a given threshold.}}\\ \hline
{\small \textit{ATE19}} & \multicolumn{1}{|l|}{{\small The measured
multicopter's distance from the base is not less than a given threshold.}%
}\\ \hline
{\small \textit{ATE20}} & \multicolumn{1}{|l|}{{\small The measured
multicopter's throttle is less than a given threshold over a time horizon.}%
}\\ \hline
{\small \textit{ATE21}} & \multicolumn{1}{|l|}{{\small Other throttle
situation.}}\\ \hline \hline
\end{tabular}
\
\]

\end{center}

Here, note that this paper assumes the health check of equipment above can be
performed by effective fault diagnosis and health evaluation methods. For
simplified presentation, the statements of \textquotedblleft check result
of\textquotedblright \ and \textquotedblleft measured\textquotedblright \ are
omitted in the subsequent sections.

\textbf{Remark 1.} MCEs are defined to guarantee the controllability of the
plant, because supervisory control restricts the behavior of a plant such that
the given control specifications are fulfilled and as much as possible never
violated, by enabling or disabling controllable events in the plant. According
to safety requirements, the user declares which mode the multicopter should
enter. This leads to the definition of controllable events related to mode
transitions, namely MCEs.

\section{Failsafe Mechanism Design}

The failsafe mechanism uses multicopter modes and switch conditions among them
to make multicopters satisfy the user's safety requirements. In this section,
functional requirements are used to model a multicopter plant automaton with
defined multicopter modes and events. Then, from the safety requirements,
multiple control specifications are represented by automata. These control
specifications should indicate the preferable failsafe measures consistent
with the textually described safety requirements. After the plant and control
specifications have been obtained, the supervisor is synthesized by using
monolithic supervisory control.

\subsection{Multicopter plant modeling}

\subsubsection{Modeling principles}

\begin{figure}[h]
\begin{center}
\includegraphics[
width=16 cm]{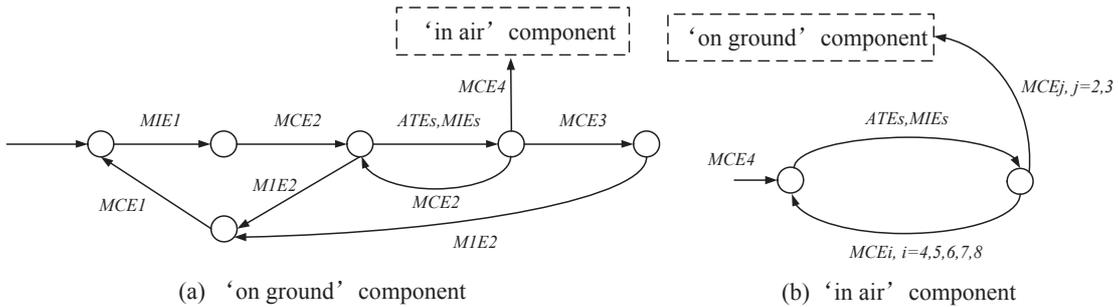}
\end{center}
\caption{Schematic diagram of the multicopter plant}%
\label{diagram_of_plant}%
\end{figure}

Modeling the multicopter plant is to mathematically describe what behavior the
multicopter is able to perform with an automaton transformed from functional
requirements. In this paper, the modeling principles of the multicopter plant
include: i) modeling from a simple schematic diagram to a comprehensive
automaton model; 2) modeling from the `on ground' component to the `in air'
component; 3) events of each transition modeled mutually exclusively. Figure
\ref{diagram_of_plant} depicts a schematic diagram of the `on ground'
component and `in air' component of the multicopter plant, respectively. The
schematic diagram lists all modes which the multicopter possibly enters, after
a series of MIEs and ATEs occur.

\subsubsection{Model details}

\begin{figure}[h]
\begin{center}
\includegraphics[
width=17 cm]{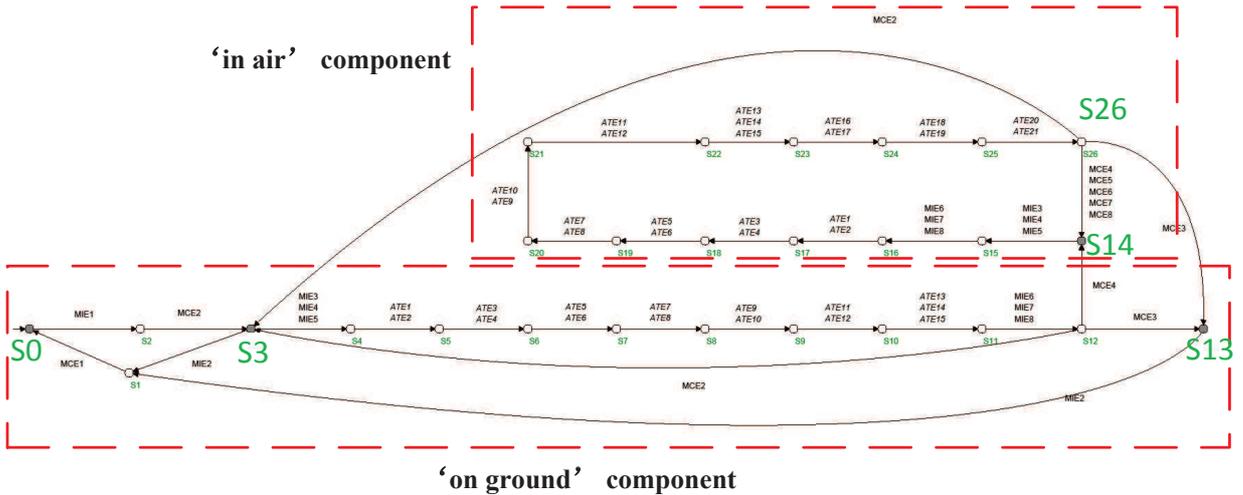}
\end{center}
\caption{Automaton model of \textit{Plant. }In this plant, the following
functions are described in the automaton model: i) if the power is turned on
(\textit{MIE1} occurs), the multicopter enters STANDBY MODE (\textit{MCE2}
occurs); ii) in STANDBY MODE, if the power is turned off (\textit{MIE2}
occurs), the multicopter enters POWER OFF MODE (\textit{MCE1} occurs); iii) in
STANDBY MODE, according to remote pilot's operation (\textit{MIE3}%
-\textit{MIE8}) and the health status of onboard equipment (\textit{ATE1}%
-\textit{ATE15}), the multicopter may either enter LOITER MODE (\textit{MCE4}
occurs), GROUND-ERROR MODE (\textit{MCE3} occurs), or stay in STANDBY MODE
(\textit{MCE2} occurs); iv) in LOITER MODE, according to remote pilot's
operation (\textit{MIE3}-\textit{MIE8}) the health status of onboard equipment
(\textit{ATE1}-\textit{ATE15}), and the multicopter status (\textit{ATE16}%
-\textit{ATE21}), the multicopter can switch among LOITER MODE (\textit{MCE4}
occurs), ALTITUDE-HOLD MODE (\textit{MCE5} occurs), STABILIZE MODE
(\textit{MCE6} occurs), RTL MODE (\textit{MCE7} occurs) and AL MODE
(\textit{MCE8} occurs); v) the multicopter can also be manually or
automatically disarmed, and enter STANDBY MODE (\textit{MCE2} occurs) or
GROUND-ERROR MODE (\textit{MCE3} occurs).}%
\label{plant}%
\end{figure}

By extending the above schematic diagrams with detailed events and
transitions, the plant is described by an automaton as shown in Figure
\ref{plant}. It describes the basic function of a multicopter. Specifically,
\textit{Plant} contains 27 states (S$_{0}$-S$_{26}$), 37 events and 63
transitions. Here, the states S$_{0},$S$_{3},$S$_{13},$S$_{14}$ are marked as
accepting states. The state S$_{0}$ represents POWER OFF MODE; the state
S$_{3}$ represents STANDBY MODE; the state S$_{13}$ represents GROUND-ERROR
MODE; the state S$_{14}$ integrates other multicopter modes. \textit{Plant}
can be divided into two parts: one (consists of states S$_{0}$-S$_{14}$ and
transitions among them) describes the multicopter behavior on the ground (`on
ground' component), and the other one (consists of states S$_{3}$,S$_{12}%
$-S$_{26}$ and transitions among them) describes the behavior during flight
(`in air' component). These correspond to the schematic diagram shown in
Figure \ref{diagram_of_plant}.

\subsection{Control specification design}

\subsubsection{Modeling principle}

In this part, control specifications are designed to restrict the behavior of
\textit{Plant} according to the description of the safety requirements. In
order to obtain a correct and non-blocking supervisor, the control
specifications must cover all possible strings (enable desirable strings and
disable the others) in the plant, and the control specifications must have no
conflict among themselves.

\subsubsection{Control specification design `on ground'}

Through a study of safety requirements, it can be seen that \textit{SR1}
describes the intended failsafe measure when the multicopter is on the ground.
In other words, \textit{SR1} restricts what action the user wants the
multicopter to perform under specific situations when it is on the ground.
Thus, we design a control specification\ to cover all possible strings in the
`on ground' component of \textit{Plant}. The requirements given in Tables 1-3
are different from the designed specifications. The former is textually and
informally described, whereas, based on which, the latter is designed formally
described in form of automaton. Several requirements may be described by one
specification, or one requirement may be described by several specifications.

\begin{figure}[h]
\begin{center}
\includegraphics[
width=16 cm]{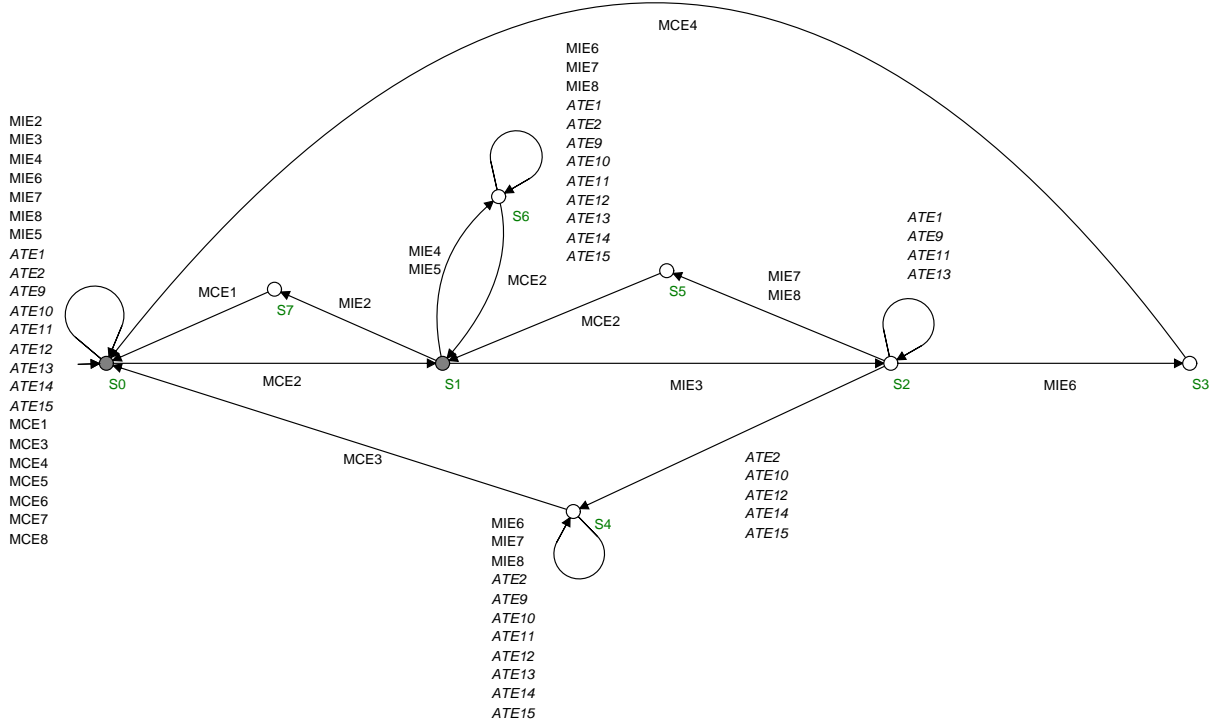}
\end{center}
\caption{Automaton model of \textit{Specification 1. }In \textit{Specification
1}, the multicopter is first in STANDBY MODE (\textit{MCE2} occurs). In this
case, when the remote pilot executes an arm action (\textit{MIE3} occurs), if
the INS and propulsors are both healthy (\textit{ATE1} and \textit{ATE9}
occur), the connection to RC transmitter is normal (\textit{ATE11} occurs),
the battery's capacity is adequate (\textit{ATE13} occurs), and the flight
mode switch is on the position of \textquotedblleft normal
flight\textquotedblright \ (\textit{MIE6} occurs), then the multicopter can be
successfully armed, and enter LOITER MODE (\textit{MCE4} occurs). Otherwise,
if the remote pilot does not execute an arm action (\textit{MIE4} or
\textit{MIE5} occurs), or the flight mode switch is not on the position of
\textquotedblleft normal flight\textquotedblright \ (\textit{MIE7} or
\textit{MIE8} occurs), the multicopter stays in STANDBY MODE (\textit{MCE2}
occurs); if one of the related component is unhealthy (\textit{ATE2},
\textit{ATE10}, \textit{ATE12}, \textit{ATE14} or \textit{ATE15} occurs), the
multicopter enters GROUND-ERROR MODE (\textit{MCE3} occurs). Also, the remote
pilot can directly turn off the power (\textit{MIE2} occurs), and the
multicopter enters POWER OFF mode (\textit{MCE1} occurs).}%
\label{Spec1}%
\end{figure}

In safety requirement \textit{SR1}, the user lists the required conditions for
a successful arm. In order to model it with an automaton, the key is to split
the branches in the `on ground' component of \textit{Plant}, and enable only
one mode which the user expects the multicopter to switch to. Following this
principle, a control specification named \textit{Specification 1} is designed
as shown in Figure \ref{Spec1}. It contains 8 states (S$_{0}$-S$_{7}$), 24
events and 68 transitions. Here, the states S$_{0},$S$_{1}$ are marked as
accepting states. The state S$_{1}$ represents STANDBY MODE, and the state
S$_{0}$ integrates other multicopter modes. Here, two points need to be noted:

i) The selfloops on the state S$_{0},$S$_{4},$S$_{6}$ are used to guarantee
that the irrelevant events will not interrupt the event sequence presented in
\textit{Plant}, and not influence the occurrence of other control specifications.

ii) \textit{SR1} itself is textually and informally described.\ It does not
mention which mode the multicopter should enter, if it cannot be successfully
armed. Furthermore, it does not take all possible strings into consideration.
In this case, during the design of control specifications, it is required to
appropriately infer the user's potential intention, and add the omitted part
to guarantee that the control specification covers all possible strings in the
`on ground' component of \textit{Plant}.

\subsubsection{Control specification design `in ground'\ (Specification 7)}

\begin{figure}[ptbh]
\begin{center}
\includegraphics[
width=10.5 cm]{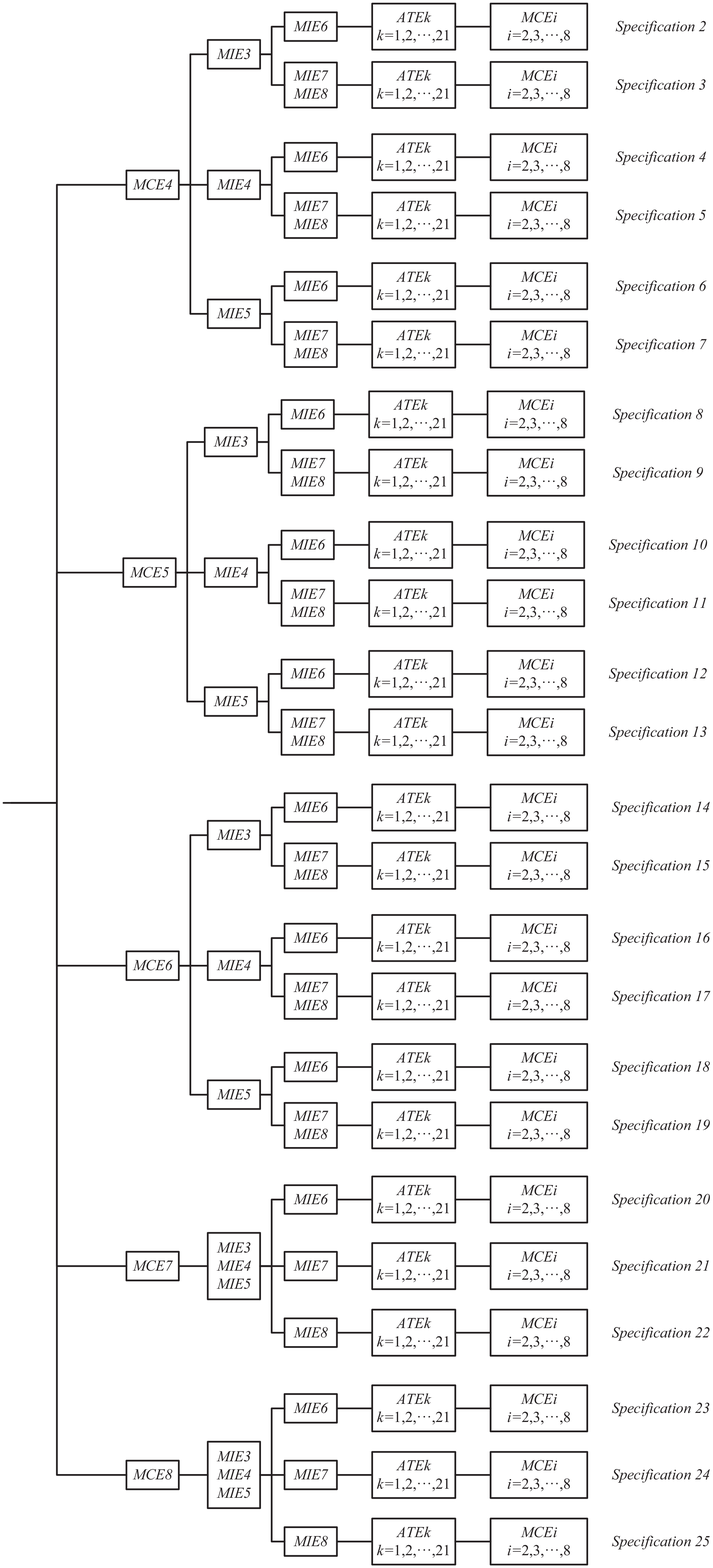}
\end{center}
\caption{Traversal relation between 24 control specifications and the
structure of the `in air' component of \textit{Plant}}%
\label{framework}%
\end{figure}

For the `in air' component of \textit{Plant}, safety requirements\textit{
SR2-SR13} restrict what action the user wants the multicopter to perform under
specific situations when it is in air. Thus, we design 24 control
specifications\ to cover all possible strings in the `in air' component of
\textit{Plant}. The traversal relation between the designed control
specifications and the structure of the `in air' component of \textit{Plant}
is shown in Figure \ref{framework}. 

\begin{figure}[h]
\begin{center}
\includegraphics[
width=11 cm]{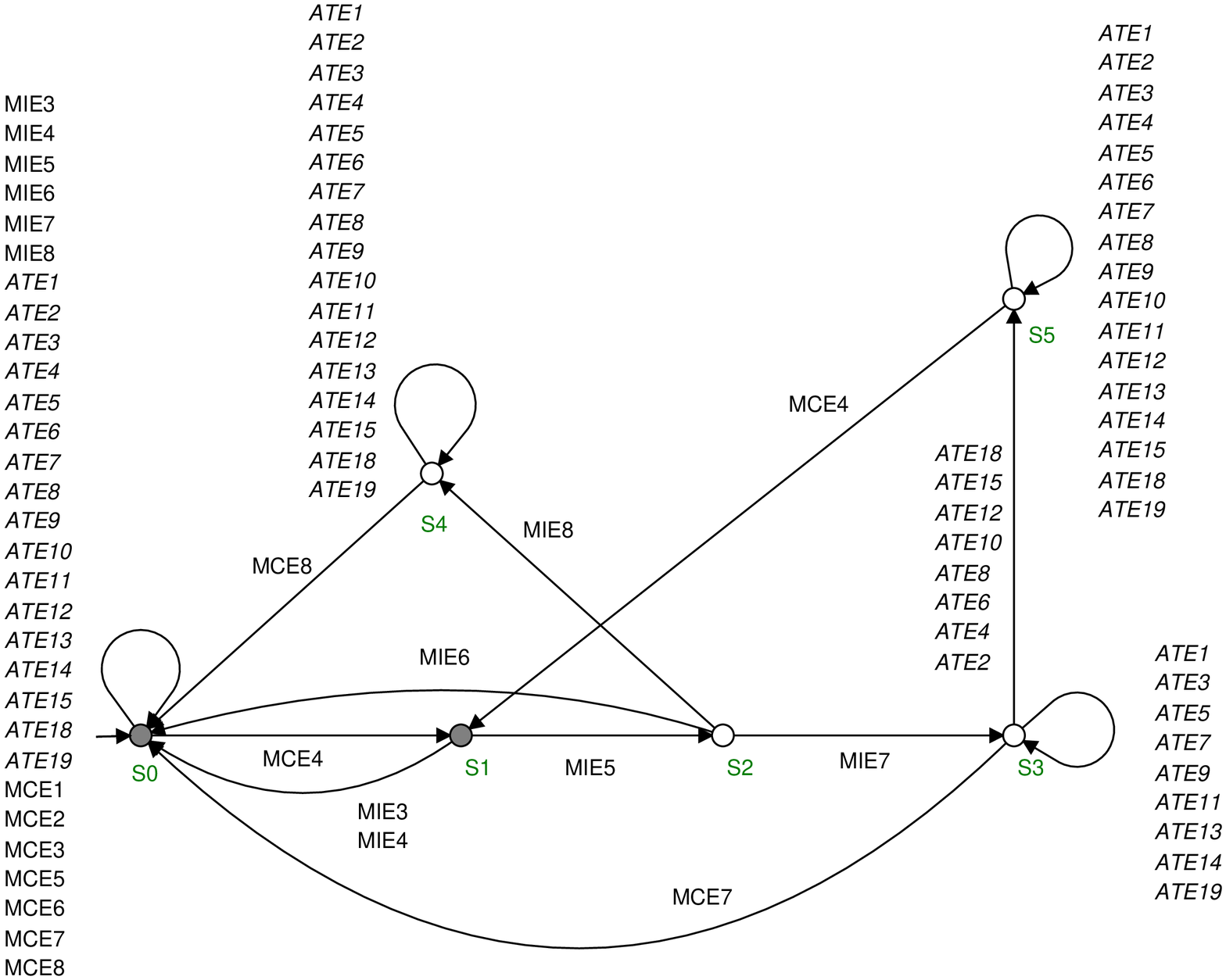}
\end{center}
\caption{Automaton model of \textit{Specification 7. Specification 7} is
triggered under the two successive conditions: i) the multicopter is in LOITER
MODE (\textit{MCE4} occurs), ii) and then the remote pilot normally
manipulates the sticks of the RC transmitter (\textit{MIE5} occurs). In this
case, when the remote pilot uses the flight mode switch to manually switch the
multicopter to RTL MODE (\textit{MIE7} occurs), if the INS, GPS, barometer,
compass, propulsors are all healthy (\textit{ATE1}, \textit{ATE3},
\textit{ATE5}, \textit{ATE7} and \textit{ATE9} occur ), the connection to the
RC transmitter is normal (\textit{ATE11} occurs), the battery's capacity is
able to support the multicopter to return to the base (\textit{ATE13} or
\textit{ATE14} occurs), and the multicopter's distance from the base is not
less than a given threshold (\textit{ATE19} occurs), then the multicopter
enters RTL MODE (\textit{MCE7} occurs); otherwise, the multicopter stays in
LOITER MODE (\textit{MCE4} occurs). Furthermore, when the remote pilot uses
the flight mode switch to manually switch the multicopter to AL MODE
(\textit{MIE8} occurs), the multicopter enters AL MODE (\textit{MCE8}
occurs).}%
\label{Spec7}%
\end{figure}

Here, because of limitation of space, we take \textit{Specification 7} as an
example to demonstrate the design of control specifications for the `in air'
component of \textit{Plant}. This control specification is obtained by
transforming \textquotedblleft safety requirements\textit{ SR7} and\textit{
SR8}\textquotedblright \ to an automaton model. As shown in Figure \ref{Spec7},
\textit{Specification 7} contains 6 states (S$_{0}$-S$_{5}$), 31 events and 91
transitions. Here, the states S$_{0},$S$_{1}$ are marked as accepting states.
The state S$_{1}$ represents LOITER MODE, and the state S$_{0}$ integrates
other multicopter modes. The details of other control specifications are
presented in the support material available in
\textit{http://rfly.buaa.edu.cn/resources}. 

\subsection{Supervisor synthesis on \textit{TCT} software}

The algorithms and operations in this part are performed on \textit{TCT}
software. In order to synthesize a supervisor by \textit{TCT} software, the
modeled multicopter plant and designed control specifications are first input.
The multicopter plant is named as \textquotedblleft \textbf{PLANT}%
\textquotedblright, and the 25 control specifications are named as
\textquotedblleft$\mathbf{E}_{j}$\textquotedblright, $j=1,2,\cdots,25$. The
input process is shown in \textit{http://rfly.buaa.edu.cn/resources}.

\subsubsection{Control specification completion}

Here, note that \textbf{PLANT}\ contains 37 events, while the number of events
in each $\mathbf{E}_{j}$ is less than 37 (i.e. the alphabet of each
$\mathbf{E}_{j}$ is different from that of \textbf{PLANT}). This is because
the given textual safety requirements only emphasize the events we are
concerned with and ignore the remaining events. For supervisory control, the
alphabet of each $\mathbf{E}_{j}$\ should be equal to the alphabet of
\textbf{PLANT}. Thus, the control specification should be completed by the
following \textit{TCT} instructions:%
\[
\mathbf{EVENTS}=\mathbf{allevents}\left(  \mathbf{PLANT}\right)
\]
where $\mathbf{EVENTS}$\ is a selfloop automaton containing all events in the
alphabet of \textbf{PLANT}. Then, for each $\mathbf{E}_{j}$, we have%
\[
\mathbf{E}_{j}=\mathbf{sync}\left(  \mathbf{E}_{j},\mathbf{EVENTS}\right)  .
\]
Here, the events present in \textbf{PLANT}\ but not in $\mathbf{E}_{j}$ are
added into $\mathbf{E}_{j}$ in form of selfloops.

\subsubsection{Supervisor synthesis}

In the monolithic supervisory control framework, all the control
specifications should be synchronized into a monolithic one. That is%
\[
\mathbf{E=sync}\left(  \mathbf{E}_{1},\mathbf{E}_{2},\cdots \mathbf{E}%
_{25}\right)  .
\]
It turns out that $\mathbf{E}$ is nonblocking, and contains 133 states and
2219 transitions. Then, a monolithic supervisor is synthesized by%
\[
\mathbf{S}=\mathbf{supcon}\left(  \mathbf{PLANT},\mathbf{E}\right)  .
\]
The obtained supervisor is the expected failsafe mechanism. It contains 784
states, 37 events and 1554 transitions. There are 8 accepting states to be
marked, which correspond respectively to 8 multicopter modes. Besides the
monolithic supervisory control, the supervisor can also be synthesized by
decentralized supervisory control, and a supervisor reduction process can also
be carried out for an easier realization in practice. The synthesis is also
carried out in the software \textit{Supremica} with the result same to
\textit{TCT}. These source files are presented in
\textit{http://rfly.buaa.edu.cn/resources}.

\section{Examples and Discussion}

This section illustrates three examples to demonstrate some possible reasons
leading to a problematic supervisor, and gives a brief discussion about the
scope of applications and properties of the method.

\subsection{Examples}

The design of control specifications is a process to understand and
re-organize the safety requirements. If the designer synthesizes a blocking
supervisor, he must recheck the correctness of control specifications and make
modifications. Here, we illustrate three examples demonstrating the blocking
phenomenon due to inappropriate design of control specifications and
conflicting safety requirements.

\begin{figure}[h]
\begin{center}
\includegraphics[
width=13 cm]{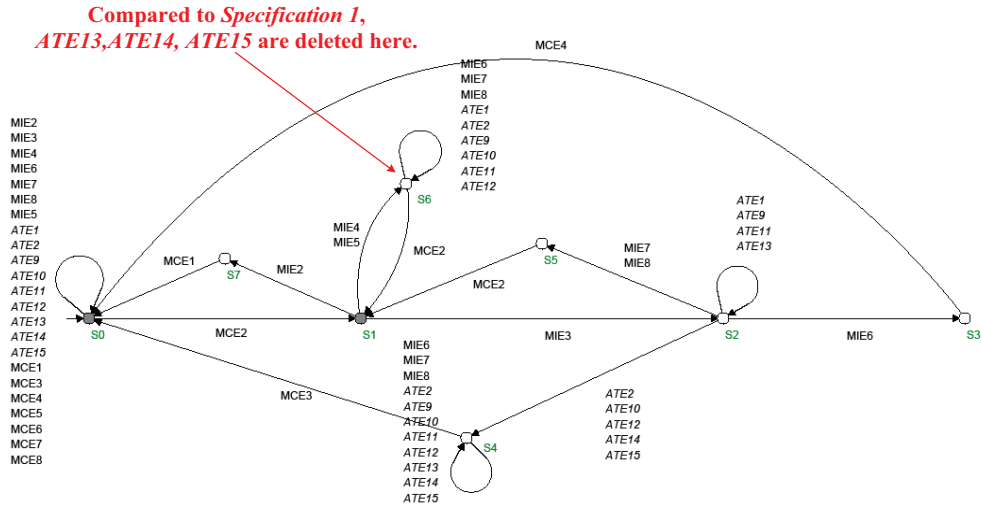}
\end{center}
\caption{Automaton model of \textit{Example 1}}%
\label{Example1}%
\end{figure}

\begin{figure}[h]
\begin{center}
\includegraphics[
width=13 cm]{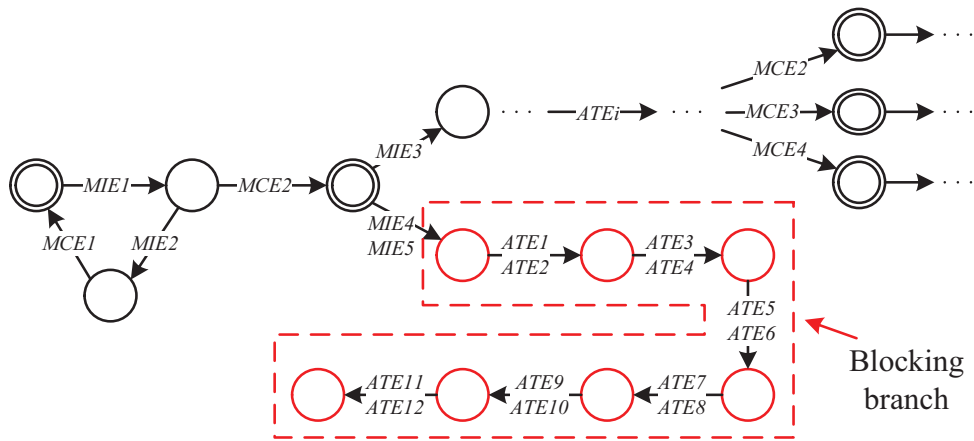}
\end{center}
\caption{Diagram of a blocking supervisor in \textit{Example 1}}%
\label{Example1_supervisor}%
\end{figure}

\textbf{Example 1}. The aim of this example is to show that missing
information in control specification may lead to a blocking supervisor. In
this example, we delete transitions \textquotedblleft S$_{6}\rightarrow
$\textit{ATE13}$\rightarrow$S$_{6}$\textquotedblright, \textquotedblleft
S$_{6}\rightarrow$\textit{ATE14}$\rightarrow$S$_{6}$\textquotedblright \ and
\textquotedblleft S$_{6}\rightarrow$\textit{ATE15}$\rightarrow$S$_{6}%
$\textquotedblright \ in \textit{Specification 1}. In this case,
\textit{Specification 1} is changed to an automaton named \textit{Example 1}
as shown in Figure \ref{Example1}. By replacing \textit{Specification 1} with
\textit{Example 1}, the supervisor is synthesized and turns out to be
blocking. The blocking branch is depicted in Figure \ref{Example1_supervisor}.
The reason is that blocking occurs owing to the missing selfloops at state
S$_{6}$ in \textit{Example 1}. The missing selfloops make the automaton
\textquotedblleft think\textquotedblright \ that events \textit{ATE13},
\textit{ATE14} and \textit{ATE15} will not occur at state S$_{6}$, while these
events should occur in \textit{Plant}. Thus, a blocking supervisor is
synthesized. This means an uncertainty as to what should occur in the blocking
point. 

\begin{figure}[h]
\begin{center}
\includegraphics[
width=13 cm]{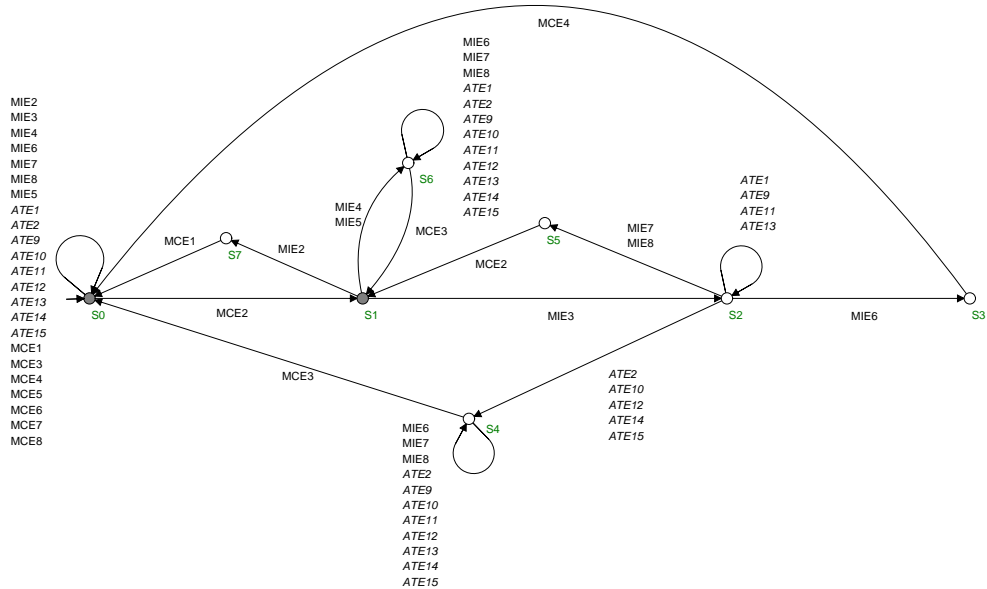}
\end{center}
\caption{Automaton model of \textit{Example 2}}%
\label{Example2}%
\end{figure}

\begin{figure}[h]
\begin{center}
\includegraphics[
width=13 cm]{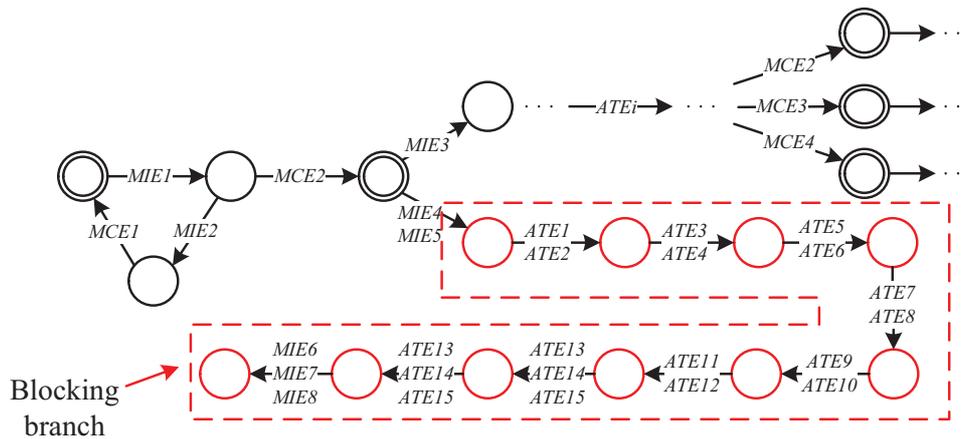}
\end{center}
\caption{Diagram of a blocking supervisor in \textit{Example 2}}%
\label{Example2_supervisor}%
\end{figure}

\textbf{Example 2}. The aim of this example is to show that conflict in
control specifications will lead to a blocking supervisor. In this example, we
replace the transition \textquotedblleft S$_{6}\rightarrow$\textit{MCE2}%
$\rightarrow$S$_{1}$\textquotedblright \ with a transition \textquotedblleft
S$_{6}\rightarrow$\textit{MCE3}$\rightarrow$S$_{1}$\textquotedblright \ in
\textit{Specification 1}. In this case, \textit{Specification 1} is changed to
an automaton named \textit{Example 2} as shown in Figure \ref{Example2}. By
adding \textit{Example 2}\ to the whole control specification, the supervisor
is synthesized and turns out to be blocking. The blocking branch is depicted
in Figure \ref{Example2_supervisor}. The reason that blocking occurs is the
conflict between \textit{Specification 1} and \textit{Example 2}.
\textit{Specification 1}\ indicates a transition \textquotedblleft
S$_{6}\rightarrow$\textit{MCE2}$\rightarrow$S$_{1}$\textquotedblright, while
\textit{Example 2}\ has a transition \textquotedblleft S$_{6}\rightarrow
$\textit{MCE3}$\rightarrow$S$_{1}$\textquotedblright. This conflict will
\textquotedblleft confuse\textquotedblright \ the supervisor, and make it
impossible to decide which transition should occur. Thus, a blocking
supervisor is synthesized. 

\begin{figure}[h]
\begin{center}
\includegraphics[
width=11 cm]{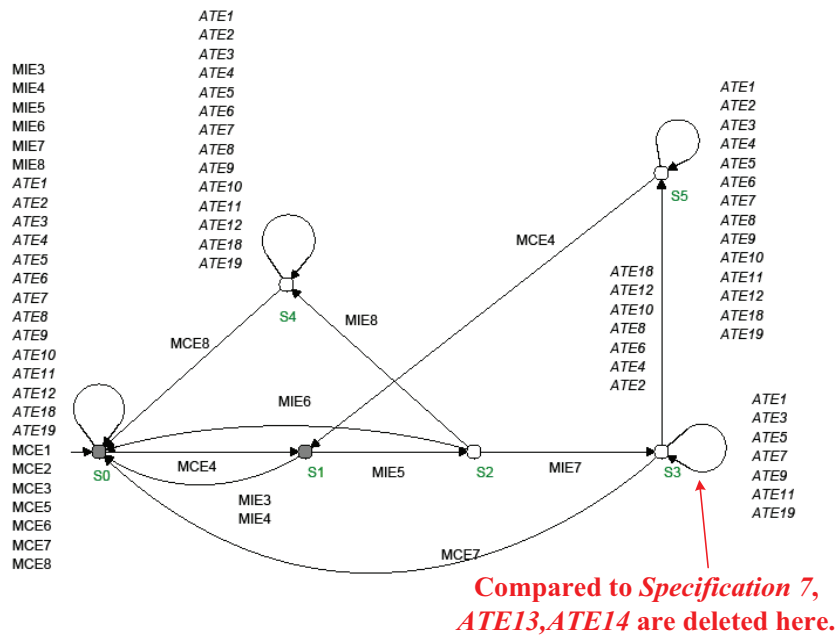}
\end{center}
\caption{Automaton model of \textit{Example 3}}%
\label{Example3}%
\end{figure}

\begin{figure}[h]
\begin{center}
\includegraphics[
width=13 cm]{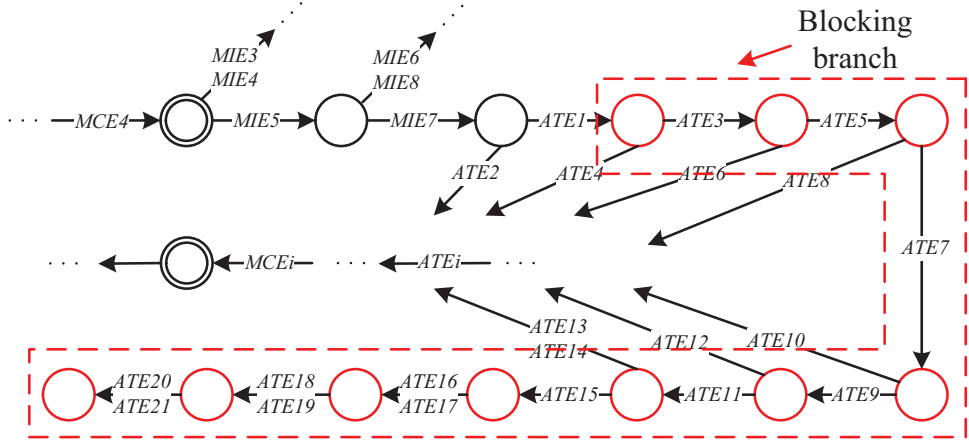}
\end{center}
\caption{Diagram of blocked supervisor in \textit{Example 3}}%
\label{Example3_supervisor}%
\end{figure}

\textbf{Example 3}. The aim of this example is to show that conflict in user
requirements will lead to a blocking supervisor. Assume we have a new safety
requirement described as follows: \textquotedblleft when the multicopter is
flying, the multicopter can be manually switched to return to the base by the
RC transmitter. This switch requires that the INS, GPS, barometer, compass and
propulsors are all healthy. Otherwise, the switch cannot
occur.\textquotedblright \ Then, this safety requirement is transformed to an
automaton as shown in Figure \ref{Example3}. By adding \textit{Example 3}\ to
the whole control specification, the supervisor is synthesized and turns out
to be blocking. The blocking branch is depicted in Figure
\ref{Example3_supervisor}. The reason that blocking appeared is the conflict
between the original \textit{SR7} and the newly presented safety requirement.
In \textit{SR7}, it indicates that \textquotedblleft this switch requires that
the INS, GPS, barometer, compass, propulsors are all healthy, and the
battery's capacity is able to support the multicopter to return to the
base\textquotedblright. However, the new safety requirement does not restrict
the condition of \textit{battery's capacity}. As in Example 2, this conflict
leads to a blocking supervisor. 

\textbf{Remark 2}. From the above examples, it can be seen that an incorrect
failsafe mechanism might be obtained during the design process due to
conflicting safety requirements or incorrect and inappropriate design of
control specifications. The mistake might be introduced inadvertently, and the
designer cannot easily detect the problem by using empirical design methods.
However, by relying on the SCT-based method, we can check the correctness of
the obtained failsafe mechanism, and make modifications if a problematic
supervisor is generated. This is a big advantage of the proposed method over
empirical design methods. Once a nonblocking supervisor is obtained, the
resulting failsafe mechanism is logically correct, and able to deal with all
relevant safety issues appearing during flight.

\subsection{Discussion}

This paper aims to study a method to guarantee the \textit{correctness} in the
design of the failsafe mechanism. Actually, correctness can be interpreted in
two different ways. On the one hand, correctness can be explained as absolute
safety, meaning that the multicopter can cope with all possible safety
problems. On the other hand, correctness is defined\ as consistency between
the obtained failsafe mechanism and safety requirements. Given a model and a
control specification for an autonomous system, synthesis approaches can
automatically generate a protocol (or strategy) for controlling the system
that satisfies or optimizes the property. This process is named as
\textquotedblleft correct-by-design\textquotedblright \  \cite{Zhang2016}. In
this domain, various formal methods and techniques, such as SCT and linear
temporal logic, are used to design control protocol of autonomous systems,
including autonomous cars \cite{Wongpiromsarn2013}, aircraft \cite{Feng2016}%
-\cite{Feng2015} and swarm robots \cite{Lopes2016}. Similar to the above
literature, in this paper, we focus on the meaning of correctness that all
requirements can be correctly satisfied. With a precise description of both
the multicopter and its correct behavior, the proposed method allows a
failsafe mechanism that guarantees the correct behavior of the system to be
automatically designed.

Here, the generated supervisor by SCT satisfies the following properties:

i) \emph{Deterministic}. This property has two aspects. First, there exists no
situation that one event triggers a transition from a single source state to
different target states in the obtained supervisor. This is a necessary
condition for a deterministic automaton. However, this situation might occur
due to man-made mistakes in an empirical failsafe mechanism design. Second,
after occurrence of MIEs and ATEs, SCT can guarantee that only one MCE is
enabled by disabling other MCEs due to deliberate design of control
specifications. In this case, after occurrence of certain MIEs and ATEs, the
mode which the multicopter should enter is deterministic.

ii) \emph{Nonconflicting} \cite[Chapter 3.6]{Wonham2009}. None of the control
specifications conflict with any others. If there exist conflicts, the
supervisor will not be successfully synthesized, because SCT cannot decide
which control specification is the user's true intention. If so, the designer
should check 1) the correctness of control specifications transforming from
user requirements; or 2) the reasonableness of the user requirements.

iii) \emph{Nonblocking}. The generated supervisor is nonblocking, which can be
interpreted that all possible strings in \textit{Plant} are considered (either
enabled or disabled) in the supervisor. If there are some strings which are
not considered in the control specifications, the marker states may not be
reached in some branches from the initial state. Then, the obtained supervisor
might be incomplete (even empty). This is because SCT cannot compute control
due to incorrect user's specifications. If so, the designer should modify the
control specifications to make them consistent.

iv) \emph{Logical correctness}. SCT is a mature and effective tool to be used
in the area of decision-making. If the plant and control specifications are
correctly modeled, the logic of the generated supervisor will correctly
satisfy user requirements without introducing man-made mistakes and bugs.

\section{Implementation and Simulation}

Based on the obtained supervisory controller generated by \textit{TCT}
software or \textit{Supremica}, an implementation method suitable for
multicopter is presented, in which the supervisory controller is transformed
into decision-making codes.

\subsection{Failsafe mechanism implementation}

On the one hand, we would want to avoid manual implementation of the
calculated supervisors, since this may introduce errors and is also difficult
for a complex case. On the other hand, we expect an easy way to generate an
Application Programming Interface (API) function with events as the input and
marked states as output so that it can be easily integrated into the existing
program in flight boards. The information required from a synthesized
supervisor is a transition matrix, which is an $m\times3$ matrix where $m$ is
the number of transitions in the synthesized supervisor (We have developed a
function to export the transition matrix based on the output file of
\textit{Supremica},\textit{ }available in
\textit{http://rfly.buaa.edu.cn/resources}). As shown in Table 8, in each row,
it consists of a source state, a destination state and a triggered event. For
example, if the multicopter is in source state $1$ and the triggered event is
$1,$ then the destination state will be $2$. By taking the synthesized
supervisor of multicopters as an example, it contains $784$ states, $37$
events and $1554$ transitions. So, the transition matrix is an $1554\times3$
matrix. In fact, we only need to consider $8$ accepting states, namely POWER
OFF MODE, STANDBY MODE, GROUND-ERROR MODE, LOITER MODE, ALTITUDE-HOLD MODE,
STABILIZE MODE, RTL MODE and AL MODE. Based on them, corresponding low-level
control actions exist. However, there exist many intermediate states in the
transition matrix ($784-8=776$ intermediate states for the considered
multicopter), to which no control actions correspond. Therefore, after one
decision period, the system must be in an accepting state. This is a major
problem we need to solve. Fortunately, this is always true.

\begin{center}
Table 8. Transition matrix%
\[%
\begin{tabular}
[c]{|c|c|c|}\hline \hline
{\small Source state} & {\small Destination state} & {\small Triggered
Event}\\ \hline \hline
${\small 1}$ & ${\small 2}$ & ${\small 1}$\\ \hline
${\small \vdots}$ & ${\small \vdots}$ & ${\small \vdots}$\\ \hline
${\small 2}$ & ${\small 3}$ & ${\small 3}$\\ \hline \hline
\end{tabular}
\]

\end{center}

In practice, the events will be detected every $0.01s$ for example, while the
decision period may be $1s$. All triggered events are collected in every
decision period. By recalling Figure \ref{plant}, since the events in every
transition are mutually exclusive, one and only one event must be triggered
for any transition. As a result, the system does not stop at intermediate
states after feeding in all detected events. For example, by recalling the `in
air' component in Figure 5, if the initial state is S14 and we collect the
events MIE5, MIE6, ATE1, ATE3, ATE5, ATE7, ATE9, ATE11, ATE13, ATE16, ATE18,
ATE20, then the system will go to S26 in \textit{Plant}. Consequently, only
one MCEi will be enabled by the autopilot according to the specifications.
Therefore, the system will stop at an accepting state finally. For our case,
the failsafe mechanism is implemented as shown in Table 9, where $\Delta>0$
represents a decision-making time interval. Actually, the high-level
decision-making should be a relatively slow process in practice. Thus, the
failsafe mechanism implementation is not synchronized with the low-level
flight control system.

\begin{center}
Table 9. Decision-making logic implementation%
\[%
\begin{tabular}
[c]{|c|l|}\hline \hline
{\small Step} & {\small Description}\\ \hline \hline
{\small 1.} & {\small Export a transition matrix from the supervisor
synthesized by TCT software or Supremica; }${\small k=0};$\\
& ${\small \Delta>0}$ {\small is a positive integer representing a
decision-making time interval; the initial state }${\small s=s}_{0}.$\\ \hline
{\small 2.} & ${\small k=k+1}$\\ \hline
{\small 3.} & {\small Detect the instruction from the RC transmitter, health
status of all considered equipments and}\\
& {\small flight status of the multicopter. If mod}$\left(  {\small k,\Delta
}\right)  =0,$ {\small go to Step 4; Otherwise, go to Step 2.}\\ \hline
{\small 4.} & {\small Generate an event set occurred in the decision-making
time interval }${\small \Delta}${\small .}\\ \hline
{\small 5.} & {\small By starting at state }$s$ {\small with the events
inputed according to the occurrence order in \textit{Plant} one by one,}\\
& {\small search the transition matrix when an event is inputed. After all the
MIEs and ATEs are inputed}\\
& {\small completely, search the transition matrix again, and only one match
will be found, where the}\\
& {\small triggered event is an MCE and the destination state is }$s_{1}%
${\small .}\\ \hline
{\small 6.} & $s=s_{1},$ {\small go to Step 2.}\\ \hline \hline
\end{tabular}
\  \
\]

\end{center}

\subsection{Simulation}

In this part, we put the failsafe mechanism into a real-time flight simulation
platform of quadcopters developed by MATLAB. Although it is realized by
MATLAB, this method is applicable to any programming language. The simulation
diagram is shown in Figure 15. This simulation contains three main functions:
i) the failsafe mechanism can determine the flight mode according to the
health check result, instruction of RC transmitter and quadcopter status; ii)
the remote pilot can fly the quadcopter through RC transmitter; iii) the
flight status of quadcopter can be visually displayed by FlightGear. Thus,
this simulation can be viewed as a semi-autonomous autopilot simulation of
quadcopters. A video of this simulation is presented in
https://www.youtube.com/watch?v=b1-K2xWbwF8\&feature=youtu.be or
http://t.cn/RXmhnu6. It contains three scenarios: i) the remote pilot manually
controls the quadcopter to arm, fly, return to launch, and land; ii) anomalies
of GPS, barometer, and INS are occurred during flight; iii) the connection of
RC transmitter is abnormal during flight.\begin{figure}[h]
\begin{center}
\includegraphics[
width=15.5 cm]{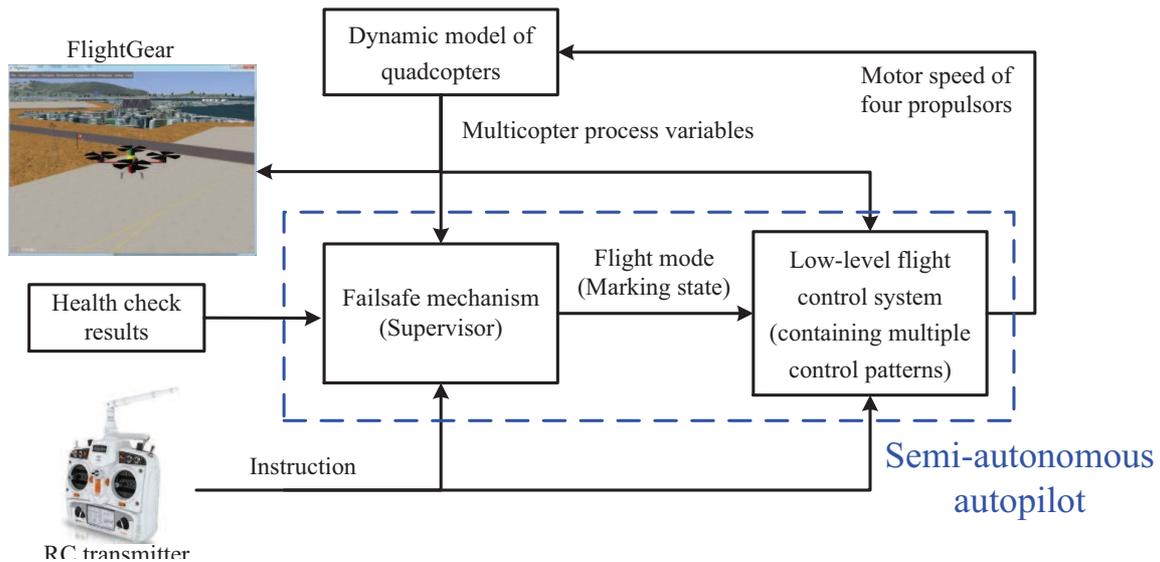}
\end{center}
\caption{Simulation diagram}%
\label{simulation_diagram}%
\end{figure}

\section{Conclusions}

This paper proposes an SCT based method to design a failsafe mechanism of
multicopters. The modeling process of system plant and control specifications
is presented in detail. The failsafe mechanism is obtained by synthesizing a
supervisor in a monolithic framework. It ignores the detailed dynamic behavior
underlying each multicopter mode. This is reasonable because the failsafe
mechanism belongs to the high-level decision-making module of a multicopter,
while the dynamic behavior can be characterized and controlled in the
low-level flight control system. Also, we discuss the meaning of correctness
and the properties of the obtained supervisor. This makes the failsafe
mechanism convincingly correct and effective, demonstrating that the proposed
method improves on purely empirical design methods. This paper deals with the
health status of multicopter components in a qualitative manner. In future
research, a quantitative health index will be added to extend the failsafe mechanism.

\end{document}